\documentstyle[mprocl]{article}

\begin{document}

\def\be{\begin{equation}}
\def\ee{\end{equation}}
\def\bea{\begin{eqnarray}}
\def\eea{\end{eqnarray}}
\def\ba#1{\begin{array}{#1}}
\def\ea{\end{array}}
\def\p{\partial}
\def\L{{\cal L}}
\def\I{{\cal I}}
\def\J{{\cal J}}
\def\dfrac#1#2{{\displaystyle\frac{#1}{#2}}}

\mathsurround=2pt
\title{\bf
A DIRECT METHOD FOR OBTAINING\\
THE DIFFERENTIAL CONSERVATION LAWS
}
\author{ A.A. CHERNITSKII }

\address{
St.Petersburg Electrotechnical University, Prof. Popov str. 5,\\
St.Petersburg 197376,  Russia ; E-mail: aa@cher.etu.spb.ru
}
\maketitle
\begin{abstract}
A direct method for obtaining the differential conservation laws
in the field theory from the
principle of stationary action is proposed.
The method is based on a variation of field functions
through small local transformation of a special kind.
The action of general theory of relativity are considered.
\end{abstract}
\par
According to the known Noether's theorem
\cite{CurantHilbert} the existence of the
differential conservation laws in the field theory is connected with
invariance of the action under continuous transformations of
dependent and independent variables.
If we have such transformation, then the Noether's theorem gives
a form of appropriate differential conservation law.
Another methods for obtaining differential conservation laws
also are used.
For example, it is of common knowledge that
the conservation law of the impulse-energy tensor can be easily
obtained from the Euler's system of equations \cite{LandauL}.
A comparatively simple method of obtaining the
conservation laws from the principle of stationary action
will be proposed here. We will derive necessary conditions of
the action stationarity (variation of the action is zero)
that can take the form of differential conservation laws.
\par
Let us consider the following action in $n$-dimensional space.
\be
S {}={} \int\limits_\Omega \L(A_\mu\,,\;B_{\nu\mu}\,,\;x^\nu)\,(dx)^n
\label{Action}
\ee
\par\hangindent=35pt \noindent \hbox to 30pt {where\hfil}
$A_\mu {}={} A_\mu(\{x^\nu\})$  are tensor components of the field,\\
$B_{\nu\mu}$ is a derivative of the function $A_\mu(\{x^\nu\})$
 by its $\nu$-th argument,\\
$\Omega$ is a region of space,
$(dx)^n {}\equiv{} dx^1 dx^2...dx^n$.
\vskip 0.5em

As it is well known, direct variation of the functions
$A_\mu(x)$ generate a Euler's system of equations as
condition of the stationary action (\ref{Action}).
However, it is possible to obtain a variation of the functions
$A_{\mu}(x)$ through a small local
transformation of some special kind. For example we can use
a small local shift of whole field configuration.
Let the functions $A_{\mu}(\{x^\nu\})$ provide the stationarity
of the action $S$ (\ref{Action}) ($\delta S {}={} 0$).
Let us consider varied functions
$A_{\mu}(\{x^\nu  {}-{}  \varepsilon \,\delta a^\nu(x) \})$.
Here $\delta a^\nu (x)  {}={} 0$ on the border of region $\Omega$.
So we have the variation of the field functions with help
of a small shift of the field configuration in the space, the
shift size being function of the coordinates.
Thus the coordinate system does not vary, it is only
the functional dependence that varies.
In any fixed point of region $\Omega$ this shift generates
the variation of values of the functions $A_{\mu}(x)$ in general case.
The functions can take any values between minimum and maximum
of the stationary configuration of the field ($\delta S  {}={} 0$).
Thus, this is variation of the functions $A_{\mu}(x)$ but with
a specific way of variation. It is evident that such variation
restricts the set of the comparison functions in general case.
Thus we can derive
only necessary conditions of the stationary action with help of this
variation; that is, if $\delta S  {}={}  0$, then we have certain
conditions for the dependent variables $A_{\mu}(x)$.
\par
We shall substitute functions
$A_{\mu}(\{x^\nu  {}-{}  \varepsilon\,\delta a^\nu(x)\})$ into Lagrangian.
Now it is necessary to extract the linear terms
with $\delta a^\nu$. In the regular way, we differentiate the action by
$\varepsilon$ and we take $\varepsilon {}={} 0$ for this purpose.
Thus we have the variation of the action in the following form.
\be
\delta S {}={} \int\limits_{\Omega}
\left(f^{\mu\nu}\, B_{\rho\nu}\;
\frac{\p \delta a^\rho}{\p x^\mu} {}+{} 
\frac{\p\L}{\p A_\nu}\, B_{\rho\nu}\;\delta a^\rho {}+{}
f^{\mu\nu}\,\frac{\p B_{\mu\nu}}{\p x^\rho}\;
\delta a^\rho\right) (dx)^n
\ee
With the partial integration and taking into account
that $\delta a^\rho (x)$ are
arbitrary functions, we have the necessary condition of the stationary action
in the following form.
\be
\frac{\p}{\p x^\mu}\left(
f^{\mu\nu}\, B_{\rho\nu}\right) {}-{}
 \frac{\p\L}{\p A_\nu}\, B_{\rho\nu} {}-{}
f^{\mu\nu}\,\frac{\p B_{\mu\nu}}{\p x^\rho} {}={} 0
\label{Euler1}
\ee
If Lagrangian does not depend explicitly
on the coordinates $x^\nu$, the
relation (\ref{Euler1}) can be written in the following conservation law form.
\be
\frac{\p}{\p x^\mu}\left(f^{\mu\nu} B_{\rho\nu}
{}-{} \L\, \delta^\mu_\rho\right) {}={} 0
\ee
For space-time case this is the known conservation law of
energy-impulse tensor.
\par
In the same way we can obtain the variations of the functions
$A_\mu (\{x^\nu\})$
through small local turn about a coordinate center. For
simplicity let us consider $A_\mu (\{x^\nu\}) $ as components
of vector field.
Then, if $A_\mu (\{x^\nu\}) $ provide the stationary action,
the varied functions will have the form
$L^\rho_{.\mu} \,A_\rho (\{L^\nu_{.\delta} x^\delta \})$,
where
$L^\rho_{.\mu} {}={} \delta^\rho_{\mu} {}+{}
\varepsilon\,\delta L^\rho_{.\mu} (x)$.\break
$ \delta L^{\rho\mu} (x) {}={} - \delta L^{\mu\rho} (x)$ are turn parameters.
Also we assume that $\delta L^\rho_{.\mu}(x) {}={} 0 $ on the border.
Here the variation of the arguments of the field functions generates
a shift,
where the shift size is proportional to the distance from
the coordinate center.
Substituting the varied functions into Lagrangian, differentiating action by
$\varepsilon$ and taking $\varepsilon {}={} 0$ we obtain variation
of the action in the following form.
\bea
\label{VarAction2}
\delta S {}={} \int\limits_{\Omega}
\left(f^{\mu\nu}\, B_{\rho\nu}\, x^\beta\;
\frac{\p \delta L^\rho_{.\beta}}{\p x^\mu} {}+{}
f^{\mu\nu}\, A_\rho\;\frac{\p\delta L^\rho_{.\nu}}{\p x^\mu} {}+{}
\frac{\p\L}{\p A_\nu}\,B_{\rho\nu}\,x^\beta\;\delta L^\rho_{.\beta}
\right. {}+{} \qquad \\
 {} + {}\left.
f^{\mu\nu}\,\dfrac{\p B_{\mu\nu}}{\p x^\rho}\,
x^\beta\;\delta L^\rho_{.\beta}
\right) (dx)^n
\nonumber
\eea
When we obtained (\ref{VarAction2}) we took into account that Lagrangian
is invariant under turn transformations with constant parameters.
We use partial integration to (\ref{VarAction2}). Using the antisymmetry of
$\delta L^{\rho\mu}$ and the arbitrariness of the turn parameters
we obtain the necessary condition of the stationary action in the
following form.
\bea
\label{Euler2}
\lefteqn{\frac{\p}{\p x^\mu}
\left[f^{\mu\nu} \left(
\vphantom{f^\mu_{.\beta}\,A_\rho}
B_{\rho\nu}\,x_\beta  {}-{} B_{\beta\nu}\,x_\rho\right)
{}+{} \left(f^\mu_{.\beta}\,A_\rho
{}-{} f^\mu_{.\rho}\,A_\beta \right)\right] {}-{} }\qquad\qquad\\[5pt]
 &{}-{}
\dfrac{\p\L}{\p A_\nu} \left(
\vphantom{\dfrac{\p B_{\mu\nu}}{\p x^\beta}\,x_\rho}
B_{\rho\nu}\,x_\beta {}-{}
B_{\beta\nu}\,x_\rho\right) {}-{}
f^{\mu\nu}\left(\dfrac{\p B_{\mu\nu}}{\p x^\rho}\,x_\beta {}-{}
\dfrac{\p B_{\mu\nu}}{\p x^\beta}\,x_\rho \right)  {}={}  0
\nonumber
\eea
If Lagrangian does not depend explicitly
on the coordinates $x^\nu$,
relation (\ref{Euler2}) can be written in the
following conservation law form.
\bea
 \frac{\p}{\p x^\mu}
\left[f^{\mu\nu}\, \left(
\vphantom{f^\mu_{.\beta}\,A_\rho {}-{} f^\mu_{.\rho}\,A_\beta}
B_{\rho\nu}\,x_\beta {}-{} B_{\beta\nu}\,x_\rho\right)
{}+{} \left(f^\mu_{.\beta}\,A_\rho {}-{} f^\mu_{.\rho}\,A_\beta \right) {}-{}
\L\,\left(
\vphantom{f^\mu_{.\beta}\,A_\rho {}-{} f^\mu_{.\rho}\,A_\beta}
\delta^\mu_\rho\,x_\beta {}-{}
\delta^\mu_\beta\,x_\rho\right)\right] {}={} 0
\eea
For space-time case this is the known conservation law of
angular momentum tensor.
\par
We shall consider now the variation of field functions through
a local transformation of more general kind than the turn.
 We consider the known Lagrangian
that gives the Einstein's equations for free gravitational field
\cite{DiracPAM}.
\be
\L {}={} g^{\mu\nu}\left(\Gamma^\rho_{\mu\sigma}\,\Gamma^\sigma_{\nu\rho} {}-{}
\Gamma^\rho_{\mu\nu}\,\Gamma^\sigma_{\rho\sigma}\right) \sqrt{|g|}
\quad;
\quad \Gamma^\rho_{\mu\nu} \equiv \frac{1}{2}\,g^{\rho\gamma}\left(
B_{\nu\gamma\mu} {}+{} B_{\mu\gamma\nu} {}-{} B_{\gamma\mu\nu}
\right)
\ee
where
$B_{\nu\gamma\mu}$ is a derivative of function $g_{\gamma\mu}(\{x^\nu\})$
with its $\nu$-th argument.
\par
The action with this Lagrangian is invariant for the transformation
$x^\rho {}\to{} G^\rho_{.\mu}\,x^\mu$, where
all components of the matrix $G^\rho_{.\mu}$
are independent and constant.
Let us consider a variation of the metric tensor components
through the transformation
$g_{\mu\delta}(\{x^\nu\})\to
G^\rho_{.\mu}\,G^\gamma_{.\delta}\;g_{\rho\gamma}
(\{G^\nu_{.\sigma}\,x^\sigma\})$, where
$G^\rho_{.\mu} {}={} G^\rho_{.\mu} (x)$.
Then a variation of the action has the following form.
\bea
 \delta S {}&=&{} \int\limits_{\Omega}
\left(\L\;\delta G^\rho_{.\rho} {}+{}
f^{\mu\nu\sigma}\, B_{\rho\nu\sigma}\, x^\beta\;
\frac{\p \delta G^\rho_{.\beta}}{\p x^\mu} {}+{}
2\,f^{\mu\nu\sigma}\,
g_{\rho\sigma}\;\frac{\p\delta G^\rho_{.\nu}}{\p x^\mu} \right. {}+{}\\
{}&+&{} \left.
\frac{\p\L}{\p g_{\mu\nu}}\,B_{\rho\mu\nu}\,x^\beta
\;\delta G^\rho_{.\beta} {}+{}
f^{\mu\nu\sigma}\,\frac{\p B_{\mu\nu\sigma}}{\p x^\rho}\,
x^\beta\;\delta G^\rho_{.\beta}
\right) (dx)^n \quad;\qquad
f^{\mu\nu\sigma}\equiv \frac{\p \L}{\p B_{\mu\nu\sigma}}
\nonumber
\eea
As we see, in this case the necessary condition of the stationary action
can be written in the following conservation law form.
\bea
\frac{\p}{\p x^\mu}
\left(f^{\mu\nu\sigma}\, B_{\rho\nu\sigma}\, x^\beta {}+{}
2\,f^{\mu\beta\sigma}\, g_{\rho\sigma}
{}-{} \L\,x^\beta\,\delta^\mu_\rho \right) {}={} 0
\eea
\vskip 1.5em
\par
So we consider some transformation of dependent and independent
variables with continuous parameters. Then we derive the
necessary conditions of the stationary action,
the variation of the field functions being obtained through
this transformation bat we take its parameters as depending
on the independent variables. Thus we have the necessary conditions of
the stationary action of a special kind.
In certain cases this conditions can be written
in a differential conservation laws form.
These cases are specified by Noether's theorem but the presented
direct method more conveniently sometimes to use.


\section*{References}
\begin{thebibliography}{99}
\bibitem{CurantHilbert} R. Curant and D. Hilbert, Method der Mathematical
Physik (Verlag von Julius Springer, Berlin, 1931).
\bibitem{LandauL}
L. Landau and E. Lifshitz, The Classical Theory of
      Fields (Pergamon, Elmsford, 4th ed., 1975).
\bibitem{DiracPAM} P.A.M. Dirac, General Theory of Relativity
(John Wiley \& Sons, New York, 1975).
\end{thebibliography}
\end{document}